# Multiscale structure-property discovery via active learning in scanning tunneling microscopy.

Ganesh Narasimha[1], Dejia Kong[1,2], Paras Regmi[3], Rongying Jin[3], Zheng Gai[1], Rama Vasudevan[1*], Maxim Ziatdinov[1,4**]

[1] *Center for Nanophase Material Sciences (CNMS), Oak Ridge National Laboratory (ORNL), Oak Ridge, Tennessee, USA – 37831*
[2] *Department of Chemistry, University of Virginia, Charlottesville, VA, USA – 22903*
[3] *SmartState Center for Experimental Nanoscale Physics, Department of Physics and Astronomy, University of South Carolina, Columbia, South Carolina, USA-29208*
[4] Current affiliation: *Physical Sciences Division, Pacific Northwest National Laboratory (PNNL), Richland, Washington, USA – 99352*

*vasudevanrk@ornl.gov

**maxim.ziatdinov@pnnl.gov

**Abstract:**

Atomic arrangements and local sub-structures fundamentally influence emergent material functionalities. The local structures are conventionally probed using spatially resolved studies and the property correlations are usually deciphered by a researcher based on sequential explorations and auxiliary information, thus limiting the throughput efficiency. Here we demonstrate a Bayesian deep learning based framework that automatically correlates material structure with its electronic properties using scanning tunneling microscopy (STM) measurements in real-time. Its predictions are used to autonomously direct exploration toward regions of the sample that optimize a given material property. This autonomous method is deployed on the low-temperature ultra-high vacuum STM to understand the structure-property relationship in a europium-based semimetal, $EuZn_2As_2$, one of the promising candidates for studying the magnetism-driven topological properties. The framework employs a sparse sampling approach to efficiently construct the scalar-property space using a minimal number of measurements, about 1 – 10 % of the data required in standard hyperspectral imaging methods. We further demonstrate a target-property-guided active learning of structures within a multiscale framework. This is implemented across length scales in a hierarchical fashion for the autonomous discovery of structural origins for an observed material property. This framework offers the choice to select and derive a suitable scalar property from the spectroscopic data to steer exploration across the sample space. Our findings reveal correlations of the electronic properties unique to surface terminations, local defect density, and point defects.

# Introduction

Material structure plays a pivotal role in driving emergent functionalities across virtually all fields such as photovoltaics[1,2], battery research[3], carbon capture[4], and quantum materials[5]. A prototypical example is that different morphisms of a material exhibit drastically different physical properties. The intricate interplay between structure and properties underscores the importance of investigating these relationships to decipher the origins of material characteristics. This knowledge, in turn, facilitates the targeted engineering and processing of materials for specific applications. In the pursuit of efficiently characterizing materials for specific applications, the development of scanning probe techniques has emerged as a transformative tool across scientific disciplines[6,7]. Among these techniques, scanning tunneling microscopy (STM) stands out as a powerful tool capable of imaging both the atomic positions and electronic structures of material surfaces[8,9]. The capability to perform spectroscopic studies using STM adds another layer of depth to the materials characterization, offering atomic-level information for understanding intriguing phenomena such as superconductivity[10], phase transitions[11,12], and other electronic phenomena[13,14].

In the context of scanning probe methods, the identification of diverse structures traditionally involves a manual process, with a human researcher correlating the material properties using auxiliary information usually acquired via microscopic and spectroscopic methods. With the STM technique, grid-based current imaging tunneling spectroscopy (CITS) is employed to gain insights into physical properties such as the density of states and magnetic response of local structures, lattice variations, surface terminations, defects, or their combinations. The exploration of structures follows a sequential approach, requiring meticulous sifting across various length scales within the sample space. This step-by-step exploration, while essential for comprehensive analysis, contributes to the time-consuming nature of experimentation in STM and limits the ability to efficiently discover multiscale structure-property relationships.

In the domain of STM imaging, the discernment of atomic structures presents another formidable challenge. Some of these structures are sparsely distributed or nearly indistinguishable, turning the search process into a laborious task akin to finding a needle in a haystack. The endeavor to locate such features often involves an exhaustive search and, at times, relies on serendipitous discoveries. Once located, there remains a persistent risk of losing the region of interest be it due to the natural course of experimentation or instability-induced processes like sample or tip drift.

Advances in machine learning, especially in the past decade, have significantly influenced scientific imaging methods[15,16], leading to the development of autonomous scanning probe techniques[17-21]. Methods such as deep learning-based computer vision in STM imaging have led to intriguing developments involving chemical structure reconstruction[22], defect identification[23], tip conditioning[24], and characterization of molecular junctions[25]. Further integration into reinforcement learning frameworks has led to the realization of specialized abilities such as tip-functionalization[26] and atom manipulation[27]. However, applications of pre-trained deep learning models naturally imply the need to retrain for every new system. Even the instances that utilize transfer learning do not continuously update with the data stream.

In this article, we demonstrate a probabilistic deep learning[28] method deployed on STM for active discovery of structure-property relationship on a magnetic topological semimetal candidate, $EuZn_2As_2$[29,30]. This method combines the correlative power of the deep learning model with human intuition in designing physics-based scalars that drive the optimization[20,31,32]. The active learning approach dynamically interacts with the experimental data and the model-inference is incorporated to steer the exploration across the sample space. This offers the ability to efficiently construct the scalar space with associated uncertainties using a minimum number of sparse measurements. We observe accurate reconstruction of the scalar property space with as little as ~ 1 % of the acquisitions in comparison to conventional hyperspectral methods. Our findings reveal correlations of the electronic properties unique to surface terminations, local defect density, and point defects.

Additionally, we demonstrate a multiscale deployment of this framework to solve the inverse problem i.e., property-guided structure discovery. This multi-scale framework is implemented in a hierarchical approach for the automated discovery of structural origins for an observed material property. This allows the experiment to parse through various length scales, distinguish features, and identify atomistic structures. The autonomous workflow used for the experiments was realized by integrating the instrumental controls, on-the-fly data analysis, and active learning models.

## Results and Discussion

**Deep Kernel Learning (DKL) Experimental Workflow**

Our goal is to have an efficient way of identifying structural features that host specific physical behavior or that cause (locally) measurable changes in the system's overall behavior. A typical approach is through the educated guess of a domain operator to decide which structures "look interesting" followed by point spectroscopy measurements at those locations. An alternative approach is to capture spectra in each pixel across the entire topographic field of view and later analyze them to link structural and spectral features. However, this approach tends to be slow and inefficient as it requires measuring spectra over a sufficiently dense grid of points, even though behaviors of interest are often localized within relatively small regions. Here we demonstrate a deep learning-driven strategy for 'intelligent' sequential spectroscopic measurements toward identifying structures exhibiting a particular behavior. The key aspect is to autonomously make an informed decision under uncertainty while leveraging the information from the previously completed measurements. From the machine learning standpoint, this means that i) one cannot rely solely on the model pre-trained with a static dataset prior to the experiment as it must be updated after each measurement; ii) a model must provide a reliable uncertainty estimate. This can be achieved through a Gaussian process (GP)-based active learning and optimization[33]. However, while effective for low-dimensional problems where a target property is usually smooth across the feature space, a standalone GP may not be able to capture abstract information present in the feature-rich morphological data. This is the niche where deep learning excels, yet it falls short in providing reliable uncertainty estimates.

The solution that we adapted is to synergize deep learning with GP via deep kernel learning (DKL)[28]. The DKL method is a combination of deep neural networks (DNN) and GP regression[33]. Specifically, if $X$ are high dimensional inputs that encode structural information and $f$ is the target property, the DKL can be defined as a probabilistic model of the form

$$f \sim \mathcal{GP}(0, K(g(x), g(x`))) \qquad (1)$$

where $g$ is a DNN and $K$ is a standard GP kernel, such as the Radial Basis Function. For the sake of brevity, we assume that the observation noise is absorbed into the computation of the kernel function.

Here, the image patches from the topographic STM scan serve as the input ($X$) to the model that is trained against the experimentally measurable material property ($f$). This property is a scalar descriptor that is extracted from the spectroscopy data. Once this target property is defined, the goal is to efficiently find a subset of structures $X$ that optimizes (maximize or minimize) it. The model is updated after each new spectroscopic measurement and is used to guide the selection of the next measurement points via a pre-defined acquisition function. To avoid overfitting, we train DKL in the Bayesian setting where constant weights are replaced with prior probabilistic distributions, and a variational inference is used to learn the corresponding posteriors. Overall, by combining the correlational strength of the DNN with expert insights in devising the scalar property from spectra, the DKL method efficiently guides experimental sampling across the sample space[31,32].

The experimental workflow for automated structure-property correlation using DKL is illustrated in **Figure 1**. In the experiment, the structural input is extracted from the STM topography while the scalar property is derived from spectroscopic measurements. **Figure 1a** shows the schematic of an STM which is used to initially image a region of interest. Thereafter point-and-shoot spectroscopic measurements are performed to gather property information associated with the local structure. This information is used for active learning via the DKL. **Figure 1b** illustrates the DKL framework which combines a DNN with a GP-based regressor. Here the structural (morphological) information is incorporated as a vector input while the property-scalar is used as a target to train the DKL model. **Figure 1c** shows the constructed DKL predicted objective-mean map that is constructed from the experimental data points.

In our implementation, each experiment begins with an initial set of twenty data points randomly measured across the region following which the DKL model is deployed. At each iteration, the previous set of experimental observations serves as the training data. The DKL prediction is used to derive the acquisition function[34], which is dynamically incorporated into the STM for each successive measurement. The experiment continues until a set of iterations is complete.

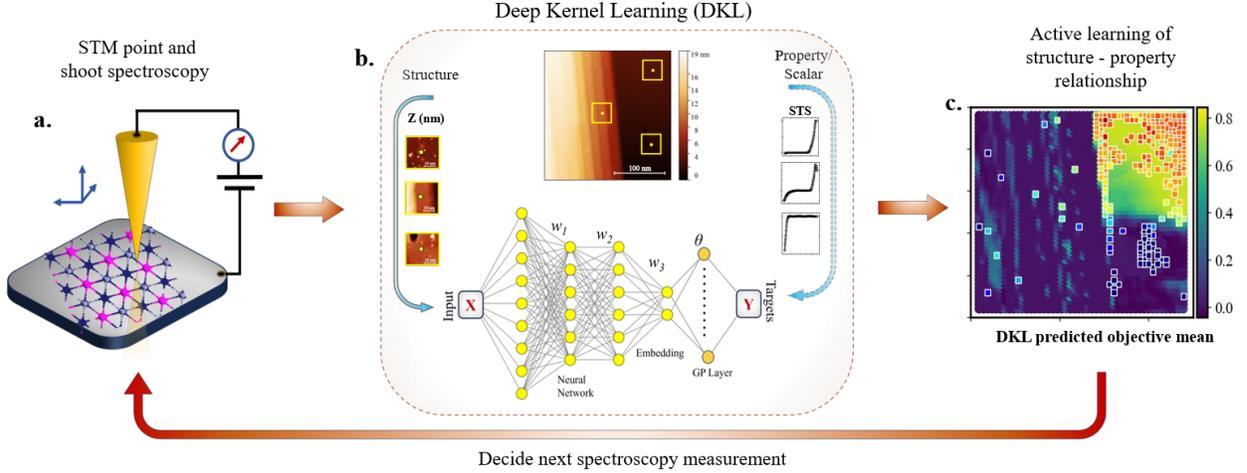

**Figure 1**: **DKL Experimental Schematic**. **(a)** Depiction of the STM experiment for imaging and spectroscopy **(b)** The Deep Kernel Learning method is a combination of the multi-layer neural network and a Gaussian Process regressor. In the active learning mode, the experimental data is used to train the model to learn the structure (morphology) – property (spectroscopy) relationship. The local morphology is the input and a physics-based scalar quantity, extracted from the spectroscopic data, is the target for training the DKL. **(c)** The DKL probabilistic prediction is used to map the structure-property relationship and a pre-selected acquisition function is used to dynamically drive the selection of the next measurement points to explore structures that maximize (or minimize) the target property.

**Characteristics of EuZn$_2$As$_2$.**

We demonstrate the utility of the DKL method for the active discovery of structure-property relationship on EuZn$_2$As$_2$, which shows anti-ferro-magnetic ordering below 19 K and strong ferromagnetic fluctuation between 19 K and 200 K[30]. At the investigated temperature (~ 77 K) the material surface exhibits a characteristic bandgap. The lattice structure of this material, shown in **Figure 2a,** is hexagonal with a Zn-As network onto which the Eu layer is superimposed. Preliminary analysis shows that the surface terminations after *in situ* low-temperature cleave could either be associated with the Zn-As network or the Eu layer. The atomically resolved image of the sample is shown in **Figure 2b**. Material morphology in a larger area, shown in **Figure 2c,** depicts predominantly two different micro-structures: planar regions and step-like regions, created during the cleavage. The atomically resolved images show that the planar regions exhibit varying degrees of defect density. We observe that the different regions and the associated features have unique spectroscopic signatures. In **Figure 2c**, the three regions marked by the squares in red, blue, and green correspond to regions of low defect density, step-like features, and high defect density, respectively. A comparison of the regions on the planar side of the sample is shown in **Figure S1**. The morphology of the three regions marked 1, 2, and 3 are shown in **Figures 2d, 2e,** and **2f**, respectively. These regions are marked by unique spectroscopic signatures. The current-voltage (*I-V*) plot obtained from the tunneling spectroscopy measurement is shown in **Figures 2g**, **2h,** and **2i**, respectively. It can be seen that the current for positive bias is higher in region 1, with intermediate values in region 2, and is significantly lower in region 3. Further, differential tunneling conductance $G = dI/dV$ for the three regions shown in **Figures 2j**, **2k,** and **2l** demonstrates a similar trend with a high value of $G$ in regions of lower defect density. The values of tunneling current and the conductance are intermediate for the step-like features shown in region 2. In the rest of the paper, we focus predominantly on the planar regions (indicated on the right side of the morphology image shown in **Figure 2c**) where we observe interesting trends in relation to the defect types and their influence on local property derived from the spectral data.

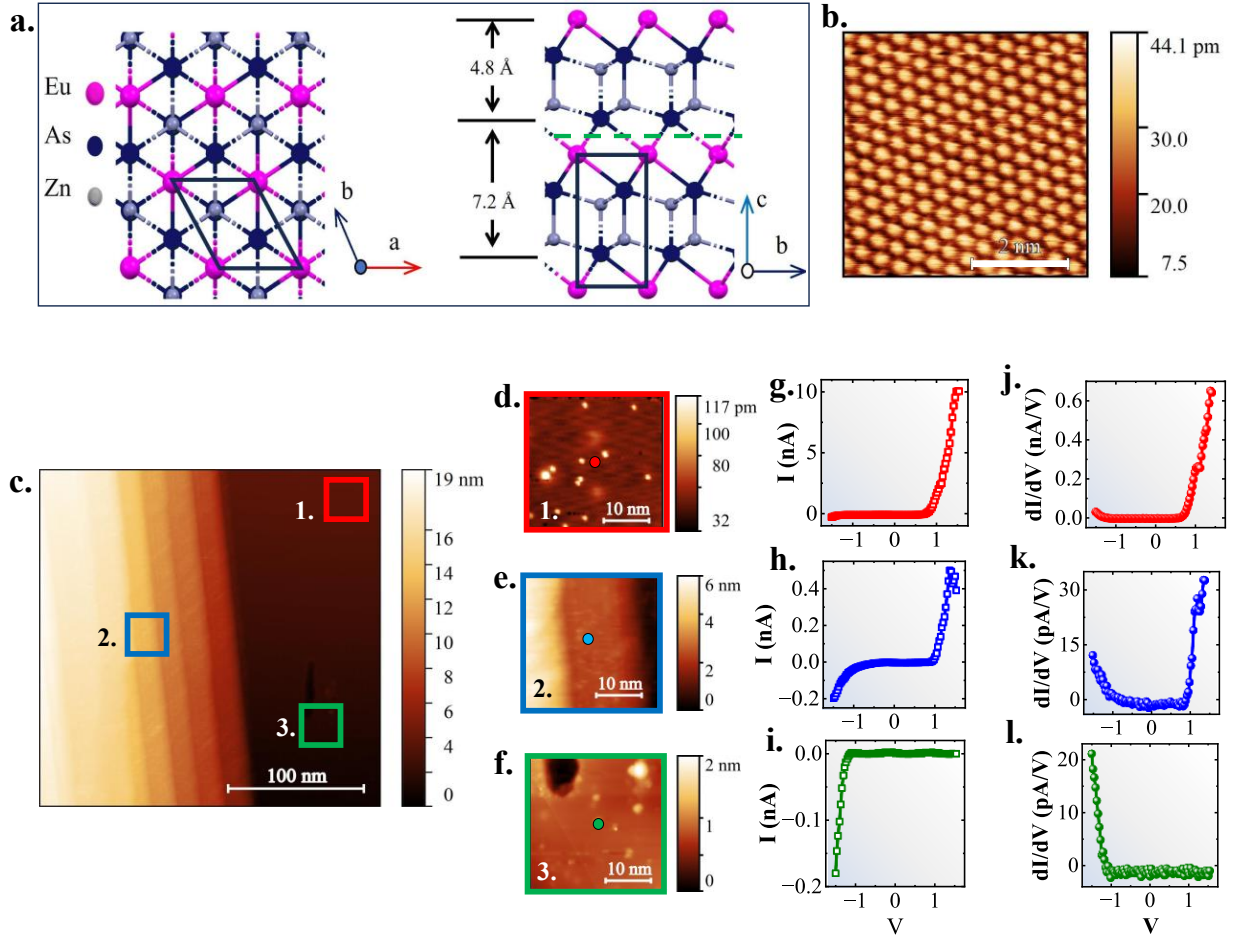

**Figure 2**: **EuZn$_2$As$_2$ material characteristics.** **(a)** The lattice structure of EuZn$_2$As$_2$ shows a hexagonal Zn-As network with an interspersed layer of Eu atoms. **(b)** shows the atomically resolved STM image of the sample surface acquired at a sample bias of – 1.5 V. The scale bar indicates 2 nm. **(c)** The characteristic surface of the material with an area of 250 nm × 250 nm shows different features. Region 1 (red square) indicates smoother regions with fewer defects. Region 2 (blue square) shows step-like features. Area 3 (green square) is indicative of a region with higher defect density. The morphology corresponding to the three regions is shown in figures **(d)**, **(e),** and **(f)**. Spectroscopic measurements are performed at the center of each region. The *I-V* data for the three regions are shown in **(g)**, **(h),** and **(i)**. The differential conductivity plots (*dI/dV* vs *V*) are shown in figures **(j)**, **(k),** and **(l)**.

**Large-area DKL implementation**

Prior to the implementation of the DKL, we performed the conventional hyperspectral measurements using CITS (with a grid density of 180 × 180 points) across the region shown in **Figure 3a** to understand the ground truth for the correlation of spectral response across the sample space. We observe that the magnitude of the current for the positive and the negative bias is different across the different regions. This is distinctly observed in the maps of the integrated current for the negative and positive bias sweep shown in **Figure S2**. Within the planar region marked by the two points A and B that correspond to the regions of low and high defect density respectively, the *I-V* plot is shown in **Figure 3b**. Point A shows a higher magnitude of the positive current in comparison to point B indicating that regions of low defect density are characterized by higher positive currents. This is clearly observed in **Figure 3c** which shows the map of the integrated current for the positive bias sweep.

With this background information, we implement the DKL to distinguish regions of high and low defect density. Note that this correlation is not readily available from the morphology information (**Figure 3a**) which was obtained at a sample bias of - 1.5 V. We implement the DKL process in the sample area shown in **Figure 3a**. Here the structural inputs are derived by fragmenting the topographic image (**Figure 3a**) into sub-image-patches while recording the location at each patch. During the experiment, spectroscopic measurements are sampled at these patch-locations. The integrated area under the positive *I-V* spectrum is considered the target scalar to be optimized (in this case, maximized) using the DKL.

**Figure 3d** shows the DKL acquisitions over 267 iterations, with morphology indicated in the background. The scatter map, which indicates the number of explorative steps, shows dense sampling in the top right region of interest. **Figure 3e** shows the scalar value map corresponding to the DKL prediction. This sparely acquired experimental data is utilized by the model to predict the scalar property across the entire field of view as shown in **Figure 3f**. This prediction correlates with the ground truth from **Figure 3c**.

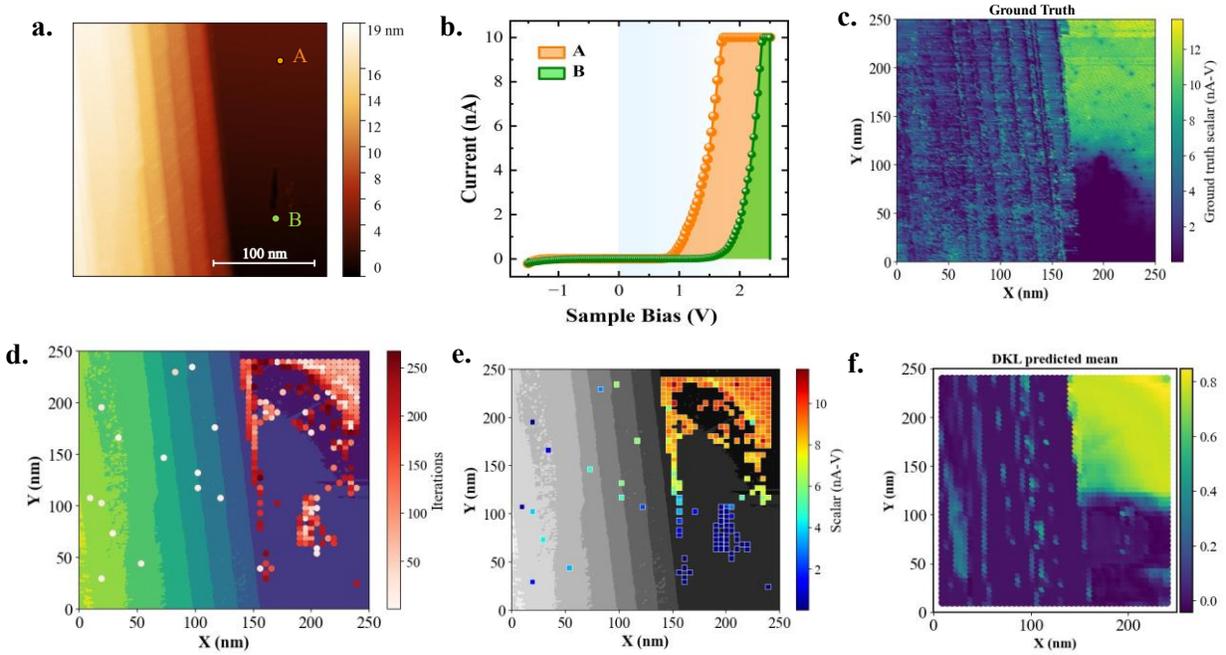

**Figure 3: Large Area DKL. (a)** Morphology of the region over an area of 250 nm × 250 nm. **(b)** Calculation of the scalar target which corresponds to the area under the *I-V* for positive states (i.e., V > 0). These values are calculated from current imaging tunneling spectroscopy (CITS) performed in the region of interest. **(c)** The target property is mapped over the entire field of view, which will serve as the "ground truth" for evaluating the DKL results. Note that the data shows distinct regions of higher and lower positive currents in the regions around points A and B. **(d)** DKL predicted measurement points over 267 iterations. The morphology is indicated in the background and the colormaps of the scatter points denote the iterations. **(e)** The measured values of the target property in the DKL predicted locations. **(f)** shows the DKL predictive mean values for objective scalar across the region of interest.

In our usage of the DKL, the optimization usually maximizes the scalar. Inverting the sign of the scalar drives the experiment to minimize the considered quantity. In this case, the minimization of the integrated positive *I-V* seeks regions of the sample space with higher defects. The DKL experiments illustrated in **Figure S3** show higher experimental acquisitions in the defect regions, especially at later iterations.

**Adsorbate defects**

In addition to a large area (250 nm × 250 nm) exploration, we implement the DKL to identify the adsorbate defects. **Figure 4a** shows a region with multiple adsorbate defects. At present, the chemical nature of these defects remains unknown. We observe that these defect structures are irregular in shape and are spectroscopically characterized by lower currents at positive sample bias. This is further reflected in lower conductivity and higher bandgap as shown in **Figure 4b**. Given this correlation, the bandgap serves as a suitable scalar descriptor.

We implement the DKL-based experimental acquisition using the bandgap (extracted from the *dI/dV* plots) as the scalar. The DKL acquisitions shown in **Figure 4c** show initial acquisitions across the region while the later iterations correspond to dense sampling on the defects. The bandgap-scalar map corresponding to the DKL predictions is shown in **Figure 4d.** We observe that the bandgap value varies across the sample. The bandgap at defects is larger than that at the surrounding defect-free region of the sample. A similar DKL method is implemented to identify these defects in a different region and the results are shown in **Figure S4**.

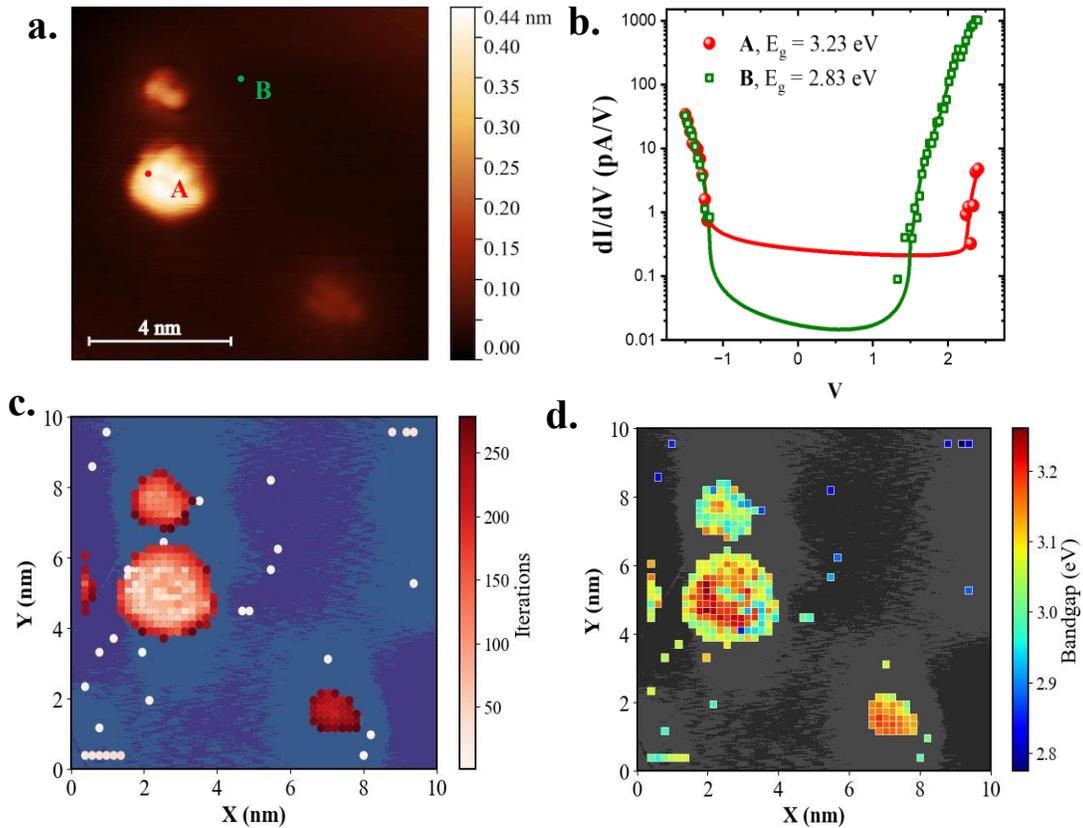

**Figure 4: DKL on adsorbate defects. (a)** The region of interest with multiple defects. Points A and B are marked on the defect and the pristine surface respectively. **(b)** tunneling spectroscopy data corresponding

to points A and B shows a higher bandgap associated with the defect region. **(c)** shows the DKL acquisition points. Here the scalar is the bandgap value to be optimized. **(d)** shows the scalar map associated with the DKL acquisitions.

**Substitution defect**

In this material, we observe a point substitution defect shown in **Figure 5a**. Our analysis suggests an atomic substitution in the Zn-As network, as depicted in the inset of **Figure 5a**. The CITS data of this defect, shown in **Figure 5b**, reveals an opposite trend with respect to the bandgap ($E_g$) energetics. The bandgap reduces at the site of these defects. A spectral comparison of the defect and the pristine surface is shown in the *dI/dV* plots in **Figure 5c**.

We implement the DKL on the defect shown in the morphology image in **Figure 5d**. Due to the inherent maximizing nature of the DKL method, we modify the *scalar = 3.1 – $E_g$* to drive the optimization to identify the lower bandgap defect regions. The DKL predictions are given by the scatter points shown in **Figure 5e** with the value of the bandgap derived from the scalar. The DKL-predicted mean of the normalized scalar across the region given in **Figure 5f** shows a correlation to the structural image in **Figure 5d**.

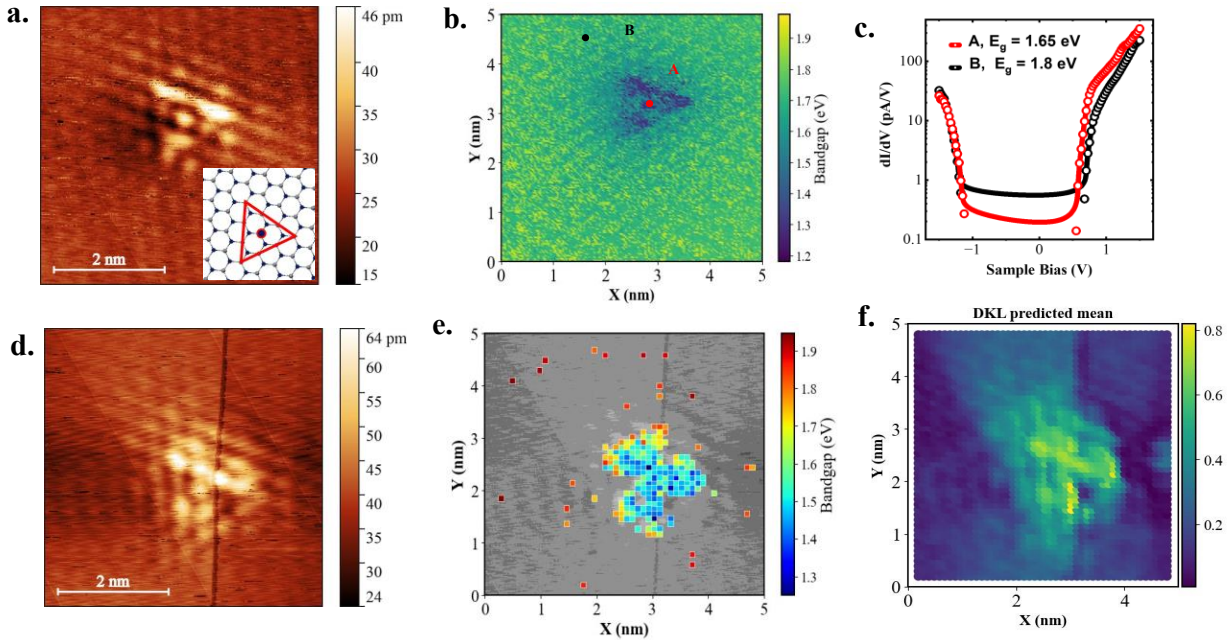

**Figure 5**: **Identification of substitution defects**. **(a)** Morphology of the substitution defect. **Inset** shows the atomistic schematic of the defect. **(b)** shows the bandgap, $E_g$ across the region, calculated from the ground truth. **(c)** shows that *dI/dV* plots of measurements on the defect and outside the defect, marked by points A and B, respectively. **(d)** Shows the image of the substitution defect that was used for the implementation of the DKL, with *scalar = 3.1 - $E_g$*. **(e)** shows the scatter points related to the DKL acquisition with the associated values of the bandgap. **(f)** shows the DKL mean prediction across the entire region of interest.

In addition to the bandgap that is lowered in the defect regions, we also observe a marginal yet consistent lowering of the valence band edge, $E_V$ (shown in **Figure S5b**). Further, a suitable DKL framework incorporating the $E_V$ value into the scalar is demonstrated as an alternative method for the identification of the substitution defect (described in **Figure S5**). This choice of designing scalars is useful to distinguish the different defect types.

**Vacancy Defect**

In addition to the substitutional defect, the DKL active-learning model is implemented on a different defect type – the vacancy defect as shown in **Figure 6a**. The atomistic schematic of the vacancy defect is shown in the inset of **Figure 6a**. The representative spectrum of points A and B is shown in **Figure 6b**. The vacancy defect shows an initial increase in conductance at lower bias which points to the presence of in-gap states. These are seen across the defect positions as shown in **Figure S6**. Further, the in-gap conductivity was observed both for positive and negative sample bias as shown in **Figure S6e**. Given this feature of the defect, we define a suitable scalar descriptor that is proportional to the in-gap conductivity, and is given by:

$$scalar = \frac{Area\ (\frac{dI}{dV}\ for\ V=[0,0.5])}{Area\ (\frac{dI}{dV}\ for\ V=[0,1.5])} = \frac{\int_0^{0.5} G dV}{\int_0^{1.5} G dV} \qquad (2)$$

The ground truth image of the scalar derived from hyperspectral CITS data is shown in **Figure 6c**.

For the DKL-guided experiments, the region of the defect is shown in **Figure 6d**. The results of the DKL acquisitions over 250 iterations given in **Figure 6e** indicate high sampling on the defect position. The corresponding DKL predicted mean across the region is shown in **Figure 6f**.

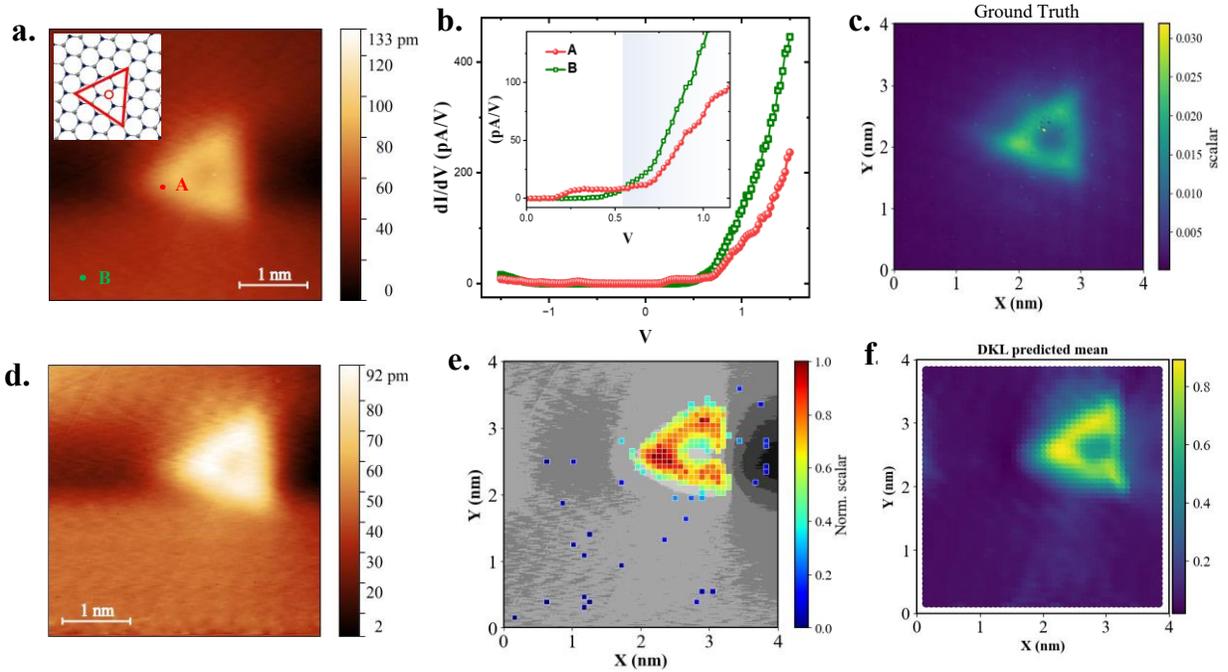

**Figure 6**: **Identification of vacancy defects**. **(a)** shows the morphology of the vacancy defect. Points A and B are marked on the defect and the pristine surface for comparison. **Inset** shows the atomistic schematic of the vacancy defect. **(b)** shows the differential conductivity plots (*dI/dV* vs *V*) for the points A and B marked on the defect and the pristine surface respectively. The magnified plot shown in the **inset** indicates higher conductivity associated with the in-gap states at the defect position. **(c)** shows the ground truth data associated with the scalar in **equation (2)**. **(d)** shows the morphology map of a vacancy defect used for DKL implementation with the scalar given by **equation (2)**. **(e)** indicates the points of DKL acquisition and the normalized scalar values. **(f)** shows the DKL-predicted mean map constructed at the end of the experiment.

Given the efficiency associated with DKL-guided experimental acquisitions, we develop a framework for the autonomous discovery of structures based on a target property. In this method, the DKL is implemented across multiple length scales to enable the exploration of atomic structures that correspond to an observed material property.

**Autonomous Exploration via DKL**

In the previous sections, each of the DKL experiments was performed at a single-length scale. Due to the resolution limitations, there is a possibility of missing structural features corresponding to multiple length scales. We can exploit DKL for learning these structures at different length scales via an iterative step-down approach. Here, we introduce a multiscale implementation of the DKL performed in the decreasing order of length scales in a cascaded fashion. This framework allows for the autonomous exploration of structures using a target property. When implemented across various length scales, this framework enables the identification of the atomistic origins of an observed property.

**Figure 7** illustrates the autonomous exploration using the multiscale DKL. Here we initially start with DKL implemented across a larger area. The next step involves the selection of a reduced area with a higher probability of "finding" the desired feature. This choice depends on the region with the maximal value of the sum of scalars corresponding to the DKL prediction. For this, we implement a sliding window of a reduced area across the whole region and calculate the sum of scalars within each placement. The position of window placement which has the highest sum of scalars is considered for subsequent implementation of DKL. This process of "zooming" continues multiple times to converge to a region of dimensions comparable to the desired atomic structure.

An example of this process is illustrated in **Figure 7**. We use this method to search the triangular vacancy defect across a large sample area of width 200 nm, shown in **Figure 7a**. The DKL utilizes the scalar given in **equation (2)** to optimize the exploration of the vacancy defect. The DKL predictions and the scalar values over 210 iterations are shown in **Figure 7e**. We then use a sliding window of a shorter dimension (100 nm) to determine the next region for exploration. The square region indicated in **Figures 7a** and **7e** corresponds to the area with the highest value of the "sum-of-scalars". This region is chosen for further implementation of DKL shown in **Figure 7b** (DKL predictions indicated in **Figure 7f**). This procedure is followed to further narrow down to an area with a width of 20 nm as shown in **Figure 7c** with the DKL-scalar map in **Figure** 7**g.** The final structure that emerges at the end of the exploration procedure is the triangular vacancy defect shown in **Figure 7d**.

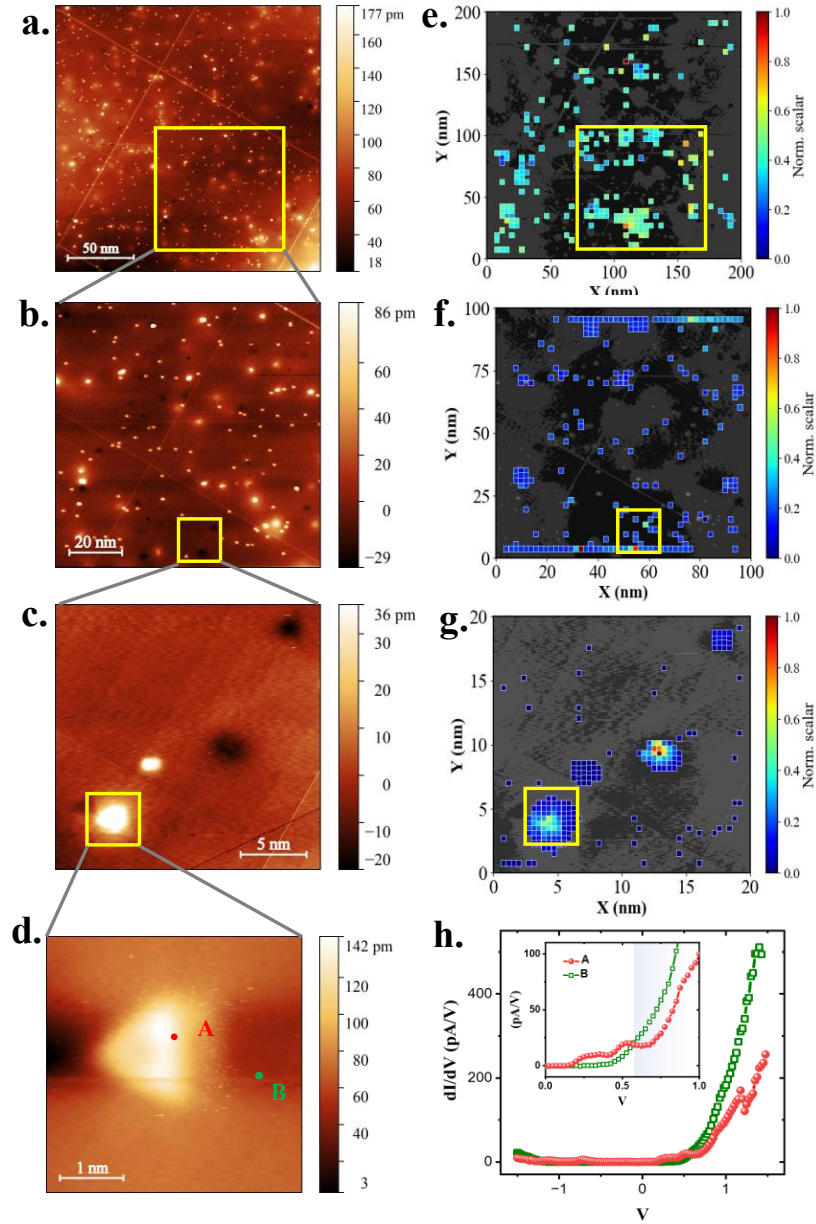

**Figure 7**: **Autonomous exploration via DKL**, to find the triangular vacancy defect using the scalar given in **equation (2)**. **(a)**, **(b),** and **(c)** show the morphology images used for successive DKL implemented in a cascaded fashion for areas with a width of 200 nm, 100, nm, and 20 nm, respectively. The yellow square in each of the images corresponds to the region with the highest value of the sum of scalars predicted within the square. This square denotes the area chosen for successive DKL implementation. **(d)** shows the final region with the identified defect obtained after 3 levels of DKL implementation. **(e)**, **(f),** and **(g)** represent the DKL prediction points and the normalized scalar values corresponding to areas shown in (a), (b), and (c) respectively. **(h)** shows the differential conductivity plots (*dI/dV* vs *V*) for the points A (on the defect)

and B (on the pristine surface) marked in (d). **Inset** shows the magnified plot indicating higher conductivity associated with the in-gap states at the defect position.

We have implemented this method to explore structures corresponding to different property scalars. In **Figure S7**, we show that maximizing the integrated positive current as the scalar drives the DKL exploration towards a planar region with no defects. On the contrary, a minimization of the positive current converges onto a large defect area, as demonstrated in **Figure S8**.

The autonomous exploration methodology can be modified and implemented to identify interesting structures and their distribution across the sample space. This method allows for the choice of the scalars utilized for the different exploration steps. It is however noted that the exploration and the convergence of a structure is probabilistic. The success of convergence depends on numerous factors such as feature size, window size of the extracted features, and resolution of acquired images. At times the target structure could be missing in a given region of the sample leading to a false convergence of the exploration framework. This can be circumvented either by enlarging the region of exploration or by distributed implementation of the multiscale-DKL framework.

Additionally, this multi-scale DKL is an example of a sequential decision-making problem, that is reminiscent of tree search algorithms. Merging recent advances in that space with DKL may provide additional opportunities for refinement. Although in this case the scalars were derived empirically by the human operator, it is possible also to envision scalars being defined on the basis of theory, to predict the probable defects and their potential impact on the local density of states. This could further be utilized to compare the structures found in the experiments with those in the simulations in an iterative framework[35].

## Conclusion

In conclusion, the DKL framework presented in this work automatically correlates material structures and electronic properties measured by STM. Our approach harnesses the power of DNNs to correlate experimental data and expert intuition to design scalar quantities that guide sample exploration toward the desired target property. The correlations and their inferences are based on sparse measurements, typically in the range of ~ 100 acquisitions. This is significantly less, in the range of 1 – 10 % of the experimental measurements using hyperspectral methods. The implementation of DKL takes about an hour, while an extensive CITS could consume ~ days. This enables real-time correlations between material structures and electronic properties, expediting the pace of discovery.

We further demonstrated a property-guided structure discovery using a multiscale DKL implemented across various length scales in a cascaded fashion for the autonomous discovery of atomistic structures. This process involves iteratively refining structural candidates based on their predicted electronic properties. The scalability and adaptability of the DKL framework make it a versatile tool for investigating topological structures, exploring emergent phenomena, characterizing impurities, or studying interfaces in heterostructures.

**Methods**

*Sample preparation*

The single crystals of $EuZn_2As_2$ were grown via the flux method using Sn. The elemental Eu (99.9% pieces, Alfa Aesar), Zn (99.8% granules, Alfa Aesar), As (99.999% powder, Alfa Aesar), and Sn (99.9% granules, Alfa Aesar) were placed into an alumina crucible with a molar ratio Eu:Zn:As:Sn = 1 : 2 : 2 : 20, and sealed in an evacuated quartz tube. The sample was heated up to 600 °C at a rate of 60 °C h$^{-1}$ and maintained at this temperature for 5 h. This was followed by warming to 1000 °C and tempering for 10 h. The sample was then slowly cooled (−3 °C h$^{-1}$) down to 600 °C with a further centrifuge in order to remove Sn flux. The resulting single crystals had a typical size of 4 mm × 2 mm × 1 mm and were stable in air. Crystal structure and basic physical properties are described in Ref.[30]

*STM experimentation*

The $EuZn_2As_2$ single crystals were cleaved on a cold stage which was cooled by liquid nitrogen under ultrahigh vacuum conditions (< 1×10$^{-10}$ torr). The samples were immediately *in situ* transferred into a pre-cooled homemade high magnetic field low temperature scanning tunneling microscope. STM/STS experiments were carried out at 77 K with base pressure lower than 1×10$^{-10}$ Torr using electrochemically etched Tungsten tips (W tip). All W tips were conditioned and checked using a clean Au (111) surface before each measurement. Topographic images were acquired in a constant current mode with a bias voltage applied to the sample. Unless specified, the topographic images were acquired at a sample bias of -1.5 V and a setpoint of 200 pA. All the spectroscopies were obtained using a lock-in amplifier with bias modulation $V_{rms}$ = 20 mV at 977 Hz. Current-Imaging-Tunneling-Spectroscopy (CITS) was performed over a grid of pixels with the same lock-in amplifier parameters.

*Deep Kernel Learning*

The data analysis related to structure-property correlation was achieved using the deep kernel learning (DKL) framework. The deep kernel consists of a deep neural network combined with a standard Gaussian Process (GP) kernel, such as a radial basis function (RBF). The DNN had three layers with the first, second, and third layers consisting of 64, 64, and 2 neurons respectively, with ReLu activation function in the first two layers. The output of the last layer was treated as inputs to the RBF kernel. The inputs for the DKL were feature vectors extracted from the STM morphology image. This was trained against a suitable property scalar that was processed from the spectroscopic data. During the active learning process, a suitable acquisition function was used to guide experimental sampling. For the experiments indicated in Figure 3 and Figure S3, the expected improvement (EI) acquisition function was used. For the rest of the experiments, the upper confidence bound acquisition function (UCB) was utilized to guide experimental exploration. In the UCB method, the $\beta$ parameter was annealed in the range of 10 – 0.001, with a 10 % reduction in every successive iteration. Open-source Python package *AtomAI* (https://github.com/pycroscopy/atomai) was used for image processing and feature extraction, while *GPax* (https://github.com/ziatdinovmax/gpax) was used for the DKL-based training and GP regression. These were integrated with LabView programs to access the STM controls.

*Bandgap Estimation*

The bandgap ($E_g$) was estimated from the conductivity vs. bias ($dI/dV$ vs $V$) spectrum. The valence ($E_V$) and the conduction band ($E_C$) edges were determined by fitting a slope at the rising side of the spectrum both for the positive and the negative bias sweep. An algorithm was implemented to consider the spectrum (i.e., $dI/dV$ vs $V$ plot) up to a certain rising threshold following which the band edge was determined by estimating the x-intercept of the linear fit. The bandgap is given by $E_g = E_V + E_C$.

# Supplementary Information

The supplementary information contains additional DKL experiments, supporting analysis of the CITS data, defect characteristics, and multiscale implementation of the DKL framework.

# Code and Data Availability

The code and the data for the DKL implemented on STM are available at: github.com/gnganesh99/DKL_on_STM.

# Acknowledgments

This work was supported by the Center for Nanophase Materials Sciences (CNMS), which is a US Department of Energy, Office of Science User Facility at Oak Ridge National Laboratory. GN acknowledges Dr. Yongtao Liu for helping with the DKL code execution.

8   Chen, C. J. *Introduction to Scanning Tunneling Microscopy Third Edition*. Vol. 69 (Oxford University Press, USA, 2021).

9   Binnig, G. & Rohrer, H. Scanning tunneling microscopy—from birth to adolescence. *reviews of modern physics* **59**, 615 (1987).

10  Fischer, Ø., Kugler, M., Maggio-Aprile, I., Berthod, C. & Renner, C. Scanning tunneling spectroscopy of high-temperature superconductors. *Reviews of Modern Physics* **79**, 353 (2007).

11  Kim, J.-J. *et al.* Observation of a phase transition from the T phase to the H phase induced by a STM tip in 1 T− TaS 2. *Physical Review B* **56**, R15573 (1997).

12  Zhang, J., Liu, J., Huang, J. L., Kim, P. & Lieber, C. M. Creation of nanocrystals through a solid-solid phase transition induced by an STM tip. *Science* **274**, 757-760 (1996).

13  Hus, S. M. *et al.* Observation of single-defect memristor in an MoS2 atomic sheet. *Nature Nanotechnology* **16**, 58-62 (2021).

14  Aradhya, S. V. & Venkataraman, L. Single-molecule junctions beyond electronic transport. *Nature nanotechnology* **8**, 399-410 (2013).

15  Kandel, S. *et al.* Demonstration of an AI-driven workflow for autonomous high-resolution scanning microscopy. *Nature Communications* **14**, 5501 (2023).

16  Moen, E. *et al.* Deep learning for cellular image analysis. *Nature methods* **16**, 1233-1246 (2019).

17  Krull, A., Hirsch, P., Rother, C., Schiffrin, A. & Krull, C. Artificial-intelligence-driven scanning probe microscopy. *Communications Physics* **3**, 54 (2020).

18  Ziatdinov, M., Ghosh, A., Wong, C. Y. & Kalinin, S. V. AtomAI framework for deep learning analysis of image and spectroscopy data in electron and scanning probe microscopy. *Nature Machine Intelligence* **4**, 1101-1112 (2022).

19  Kalinin, S. V., Ghosh, A., Vasudevan, R. & Ziatdinov, M. From atomically resolved imaging to generative and causal models. *Nature Physics* **18**, 1152-1160 (2022).

20  Liu, Y. *et al.* Learning the right channel in multimodal imaging: automated experiment in piezoresponse force microscopy. *npj Computational Materials* **9**, 34 (2023).

21  Narasimha, G., Hus, S., Biswas, A., Vasudevan, R. & Ziatdinov, M. Autonomous convergence of STM control parameters using Bayesian optimization. *APL Machine Learning* **2** (2024).

# Supplementary Information

# Multiscale structure-property discovery via active learning in scanning tunneling microscopy.

# Contents

1. **Planar region with different defect densities.**

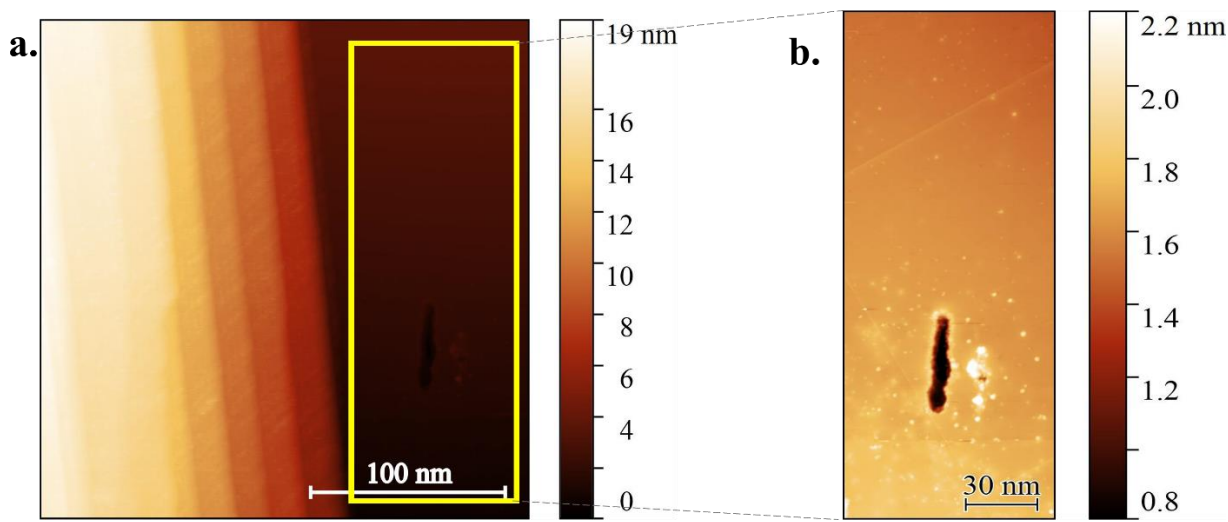

**Figure S1**: **(a)** shows the morphology image of the sample with the planar region to the right of the image. **(b)** shows the morphology of only the right planar region. The upper part of the planar regions is characterized by lower defect density.

2. **Integrated current maps for the large-area STM image**.

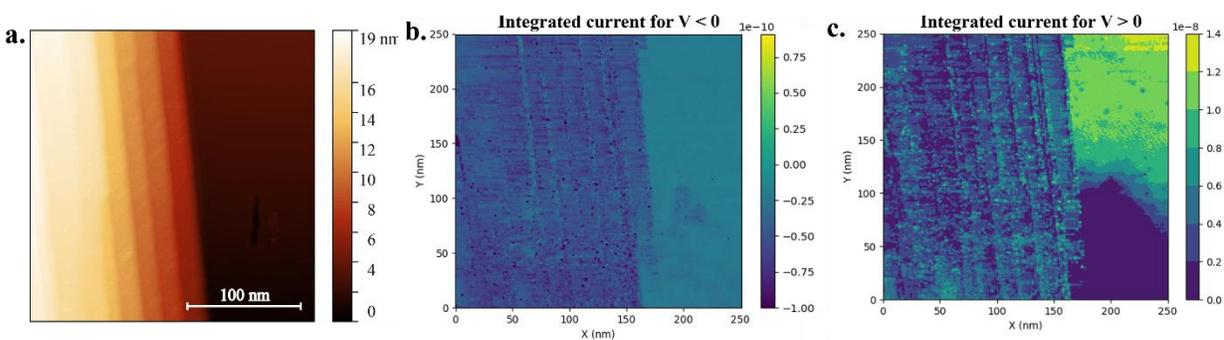

**Figure S2**: **(a)** shows the morphology of the sample across which the hyperspectral CITS imaging was performed. **(b)** shows the map of the magnitude of integrated I-V obtained for negative bias sweep, $V < 0$. **(c)** shows the map of the magnitude of integrated I-V obtained for positive bias sweep, $V > 0$.

## 3. DKL with integrated current minimization

Here we aim to drive the experimental exploration towards the region of defects that are characterized by lower magnitude of the positive current. We therefore optimize the scalar towards minimizing the integrated positive current. The scalar is defined as,

$$scalar = k - \int IdV, for\ V > 0 \qquad Eq\ (S1)$$

Here $k$ is a constant of value $k$ = 15000 pA-V

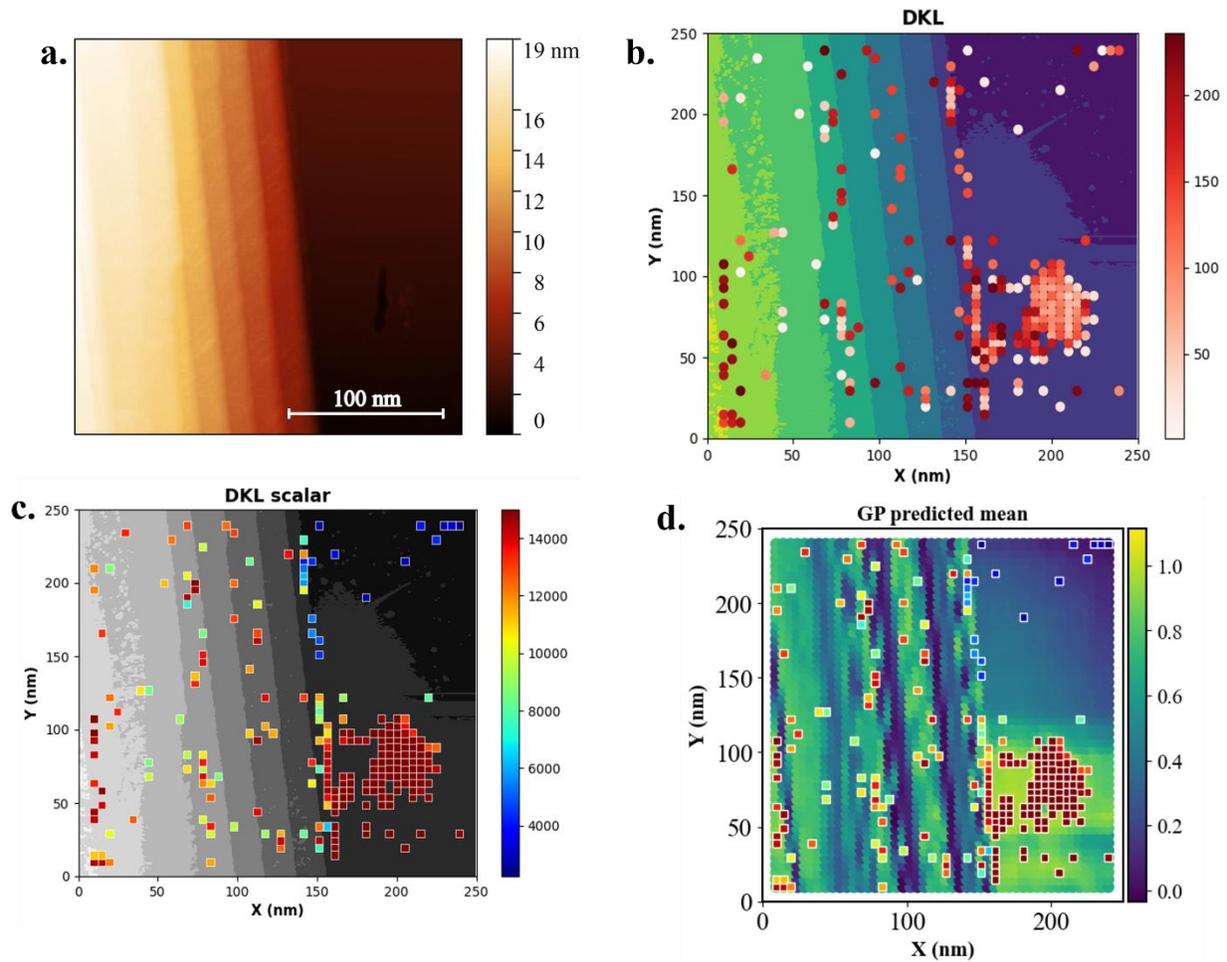

**Figure S3**: **(a)** shows the morphology of the region with an area of 250 nm × 250 nm. **(b)** shows the DKL prediction over 230 iterations with the scalar given as the (15000 - $\int IdV, for\ V > 0$). The morphology is indicated in the background and the colormaps of the scatter points denote the iterations. **(c)** shows the scalar maps corresponding to the DKL measurement points. **(d)** shows the GP mean map of the objective scalar across the region of interest. The scatter points correspond to the experimental data.

## 4. Single ad-atom defect with different bandgap values

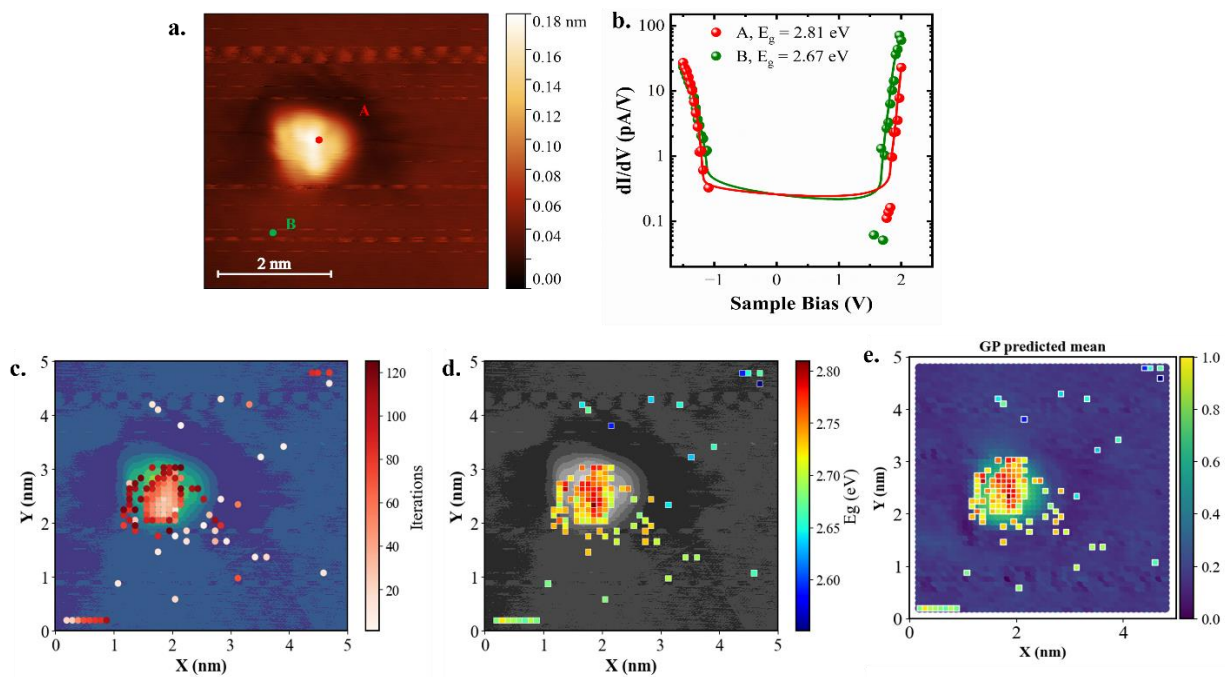

**Figure S4: (a)** The region of interest with a single defect is shown. STS is compared on the defect and outside of the defect as indicated by points A and B. STS data shown in **(b)** shows a higher bandgap on the defect region. **(c)** shows the DKL prediction with bandgap as the scalar to be maximized. **(d)** shows the scalar map in the region of interest. **(e)** shows the GP mean map of the objective bandgap scalar.

## 5. Substitution defect identification using the valence edge (*E*v) estimate.

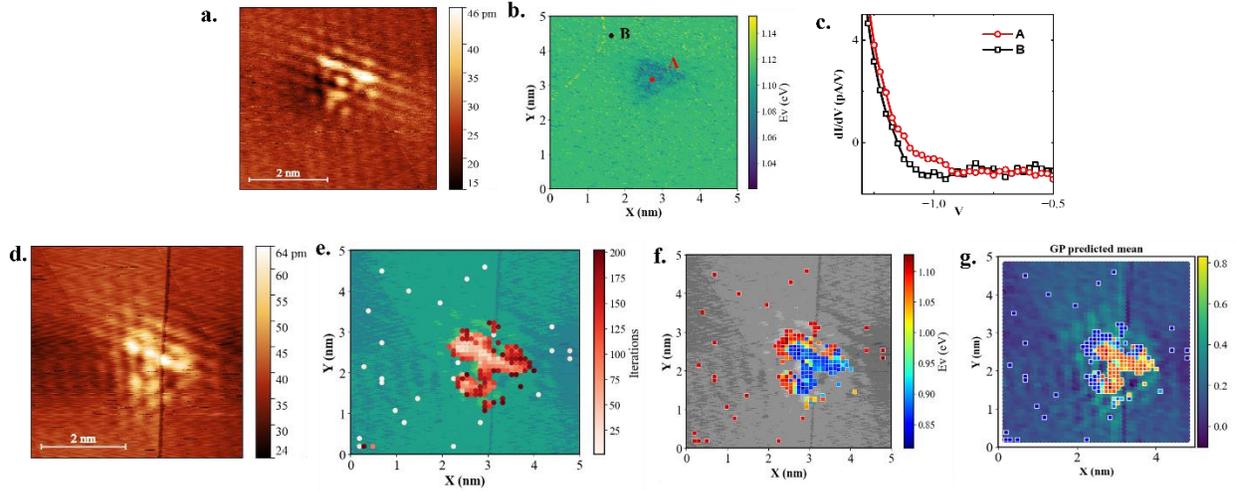

**Figure S5**: **(a)** Morphology of the substitution defect. **(b)** shows the valence band edge, $E_V$ across the region, calculated from the ground truth CITS data. **(c)** shows the *dI/dV* for V < 0 for the points marked A and B corresponding to measurements on and outside the defects, respectively. **(e)** Shows the image of the substitution defect which was used for the implementation of the DKL. The *scalar* = $2.1 - E_V$, was designed to optimize for minimized valence band offset ($E_V$) values. **(i)** shows the DKL acquisition points. **(j)** shows the *E*v-value map corresponding to the DKL points that were extracted from the scalar, and **(k)** shows the GP mean prediction map.

## 6. Vacancy defect characteristics.

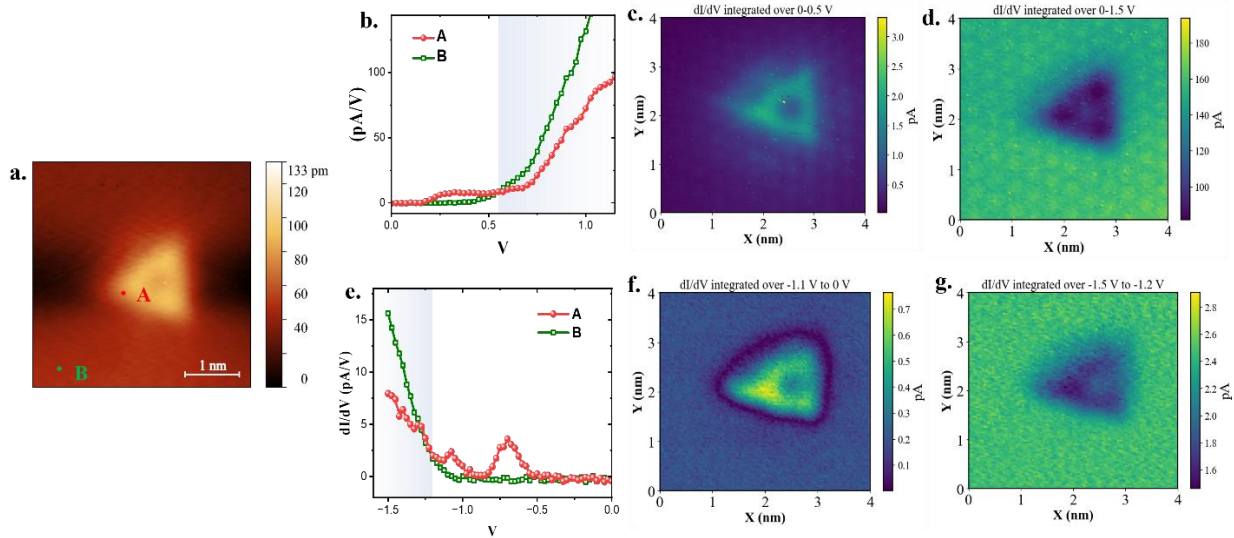

**Figure S6**: **(a)** shows the morphology of the vacancy defect. Points A and B are marked on the defect and the pristine surface for comparison. **(b)** shows the conductivity plots (dI/dV vs V) for the positive bias sweep corresponding to points A and B. **(c)** shows the map of integrated (*dI/dV* vs *V*) for V in the range [0 V, 0.5 V]. **(d)** shows the map of integrated (*dI/dV* vs *V*) for V in the range [0 V, 1.5 V]. lower panel shows the spectral analysis for the negative bias sweep. **(e)** shows the conductivity plots (*dI/dV* vs *V*) for the negative bias sweep corresponding to points A and B. **(f)** shows the map of integrated (*dI/dV* vs *V*) for V in the range [-1.1 V, 0 V]. **(g)** shows the map of integrated (*dI/dV* vs *V*) for V in the range [-1.5 V, -1.2 V].

## 7. Autonomous exploration – maximizing integrated current.

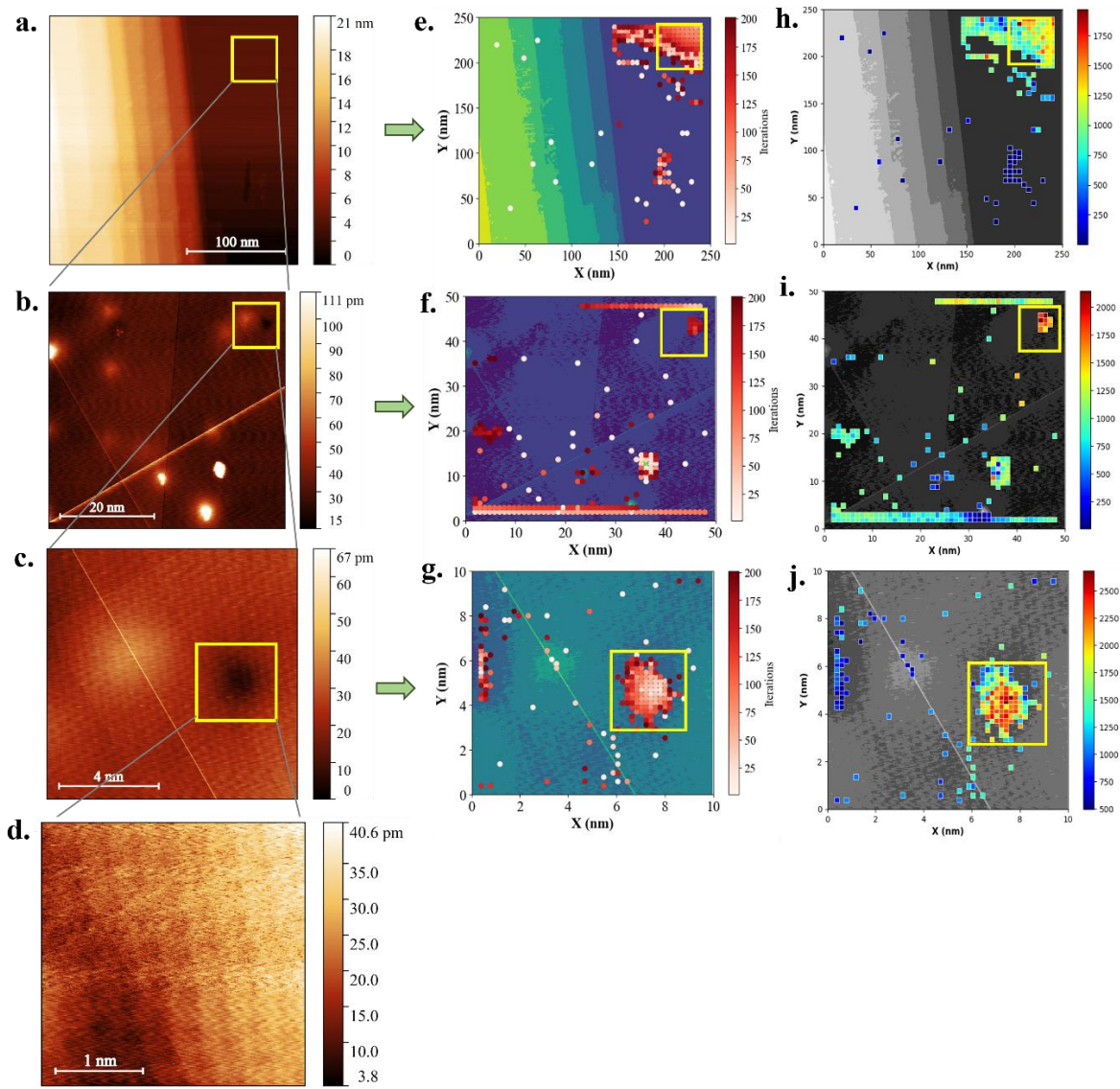

**Figure S7**: Autonomous exploration via DKL, with scalar given by the area under the positive I-V sweep. **(a)**, **(b)** and **(c)** show the morphology images used for successive DKL implemented in a cascaded fashion for areas with a width of 250 nm, 50, nm, and 10 nm, respectively. The yellow square in each of the images corresponds to the region with the highest value of the sum of scalars predicted within the square. This square denotes the area chosen for successive DKL implementation. **(d)** shows the final region with atomic resolution obtained after 3 levels of DKL implementation. **(e)**, **(f)** and **(g)** represent the DKL prediction corresponding to areas shown in (a), (b) and (c) respectively. The corresponding scalar maps are shown in **(h)**, **(i)**, and **(j)**.

## 8. Autonomous exploration – minimizing integrated current

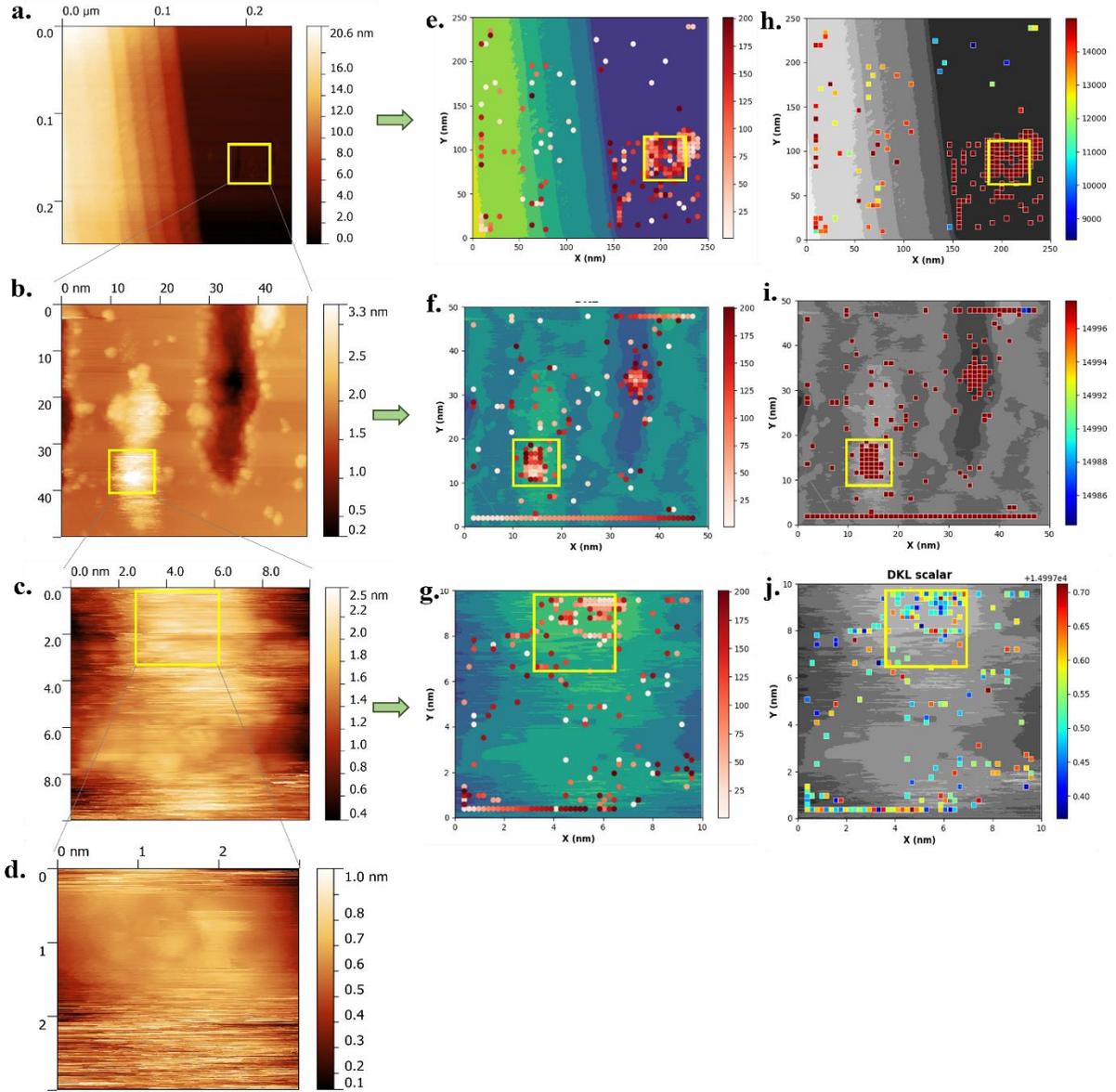

**Figure S8**: Autonomous exploration via DKL, with scalar given to minimize the area under the positive I-V sweep, using the scalar given in **Eq (S1)**. **(a)**, **(b),** and **(c)** show the morphology images used for successive DKL implemented in a cascaded fashion for areas with a width of 250 nm, 50, nm, and 10 nm, respectively. The yellow square in each of the images corresponds to the region with the highest value of the sum of scalars predicted within the square. This square denotes the area chosen for successive DKL implementation. **(d)** shows the final region of the defect obtained after 3 levels of DKL implementation. **(e)**, **(f),** and **(g)** represent the DKL prediction corresponding to areas shown in (a), (b) and (c) respectively. The corresponding scalar maps are shown in **(h)**, **(i)**, and **(j)**.